\documentclass[twocolumn,showpacs,preprintnumbers,amsmath,amssymb,superscriptaddress,aps,prl]{revtex4}
\usepackage{color,graphics}
\usepackage{verbatim}
\usepackage{amsmath}
\usepackage{graphicx}
\usepackage{bm}
\usepackage{subfigure}
\usepackage{afterpage}
\usepackage{setspace}

\newcommand{\Cm}{\rho_{\rm m}}		
\newcommand{\Cth}{\rho_{\rm thresh}}	
\newcommand{\Co}{\rho_{o}}		
\newcommand{\kon}{k_{\rm on}}	
\newcommand{\am}{\alpha} 
\newcommand{\nm}{n_{\rm c}}		
\newcommand{\lo}{L_{x}(0)}	
\newcommand{\lx}{L_{ x}}		
\newcommand{\ly}{L_{ y}}		
\newcommand{\kb}{k_{\rm B}}		
\newcommand{\Jc}{J_{\rm c}}		
\newcommand{\lc} {\ell_{c}}                  

\begin{document}
\title{Self-organized periodicity of protein clusters in growing bacteria}
\author{Hui Wang}
 \altaffiliation[Current Address:]{Department of Physics and Optical Science, University of North Carolina at Charlotte, Charlotte, North Carolina 28223}
\affiliation{Department of Physics, Clark University, Worcester, Massachusetts 01610}
\author{Ned S. Wingreen}
\affiliation{Department of Molecular Biology, Princeton University, Princeton, New Jersey 08544}
\author{Ranjan Mukhopadhyay}
 \email{ranjan@clarku.edu}
\affiliation{Department of Physics, Clark University, Worcester, Massachusetts 01610}

\date {\today}

\begin{abstract}
Chemotaxis receptors in {\it E. coli} form clusters 
at the cell poles and also laterally along the cell body, and this clustering
plays an important role in signal transduction. Recently, experiments
using fluorescence imaging have shown that, during cell growth, lateral clusters form at positions approximately periodically spaced along the cell body. In this paper, we demonstrate within a lattice model that such spatial organization could arise spontaneously from a stochastic 
nucleation mechanism. The same mechanism may explain the recent observation of periodic aggregates of misfolded proteins in {\it E. coli}.

\end{abstract}
\maketitle
     
Spatial organization of proteins is important in many cellular processes
including growth, division, movement, and establishment of polarity~\cite{Kroos03}. Over the past few years, advances in imaging techniques such as fluorescence microscopy have led to an increased appreciation of the scope and character of protein organization in cells. 
For example, 
in {\it Escherichia coli} and other bacteria, chemosensory complexes form large clusters containing thousands of receptors \cite{Zhang07}.
 Clustering of these receptors plays a crucial role in the signal integration and receptor cooperativity required for chemotaxis~\cite{Sourjik04}, {\it i.e.}  directed movement in chemical gradients. Recent work by Thiem {\it et al.}~\cite{ThiemPositioningofchemosensoryEJ07, Sourjik08}   demonstrated that clusters of chemotaxis receptors are approximately periodically positioned along the cell wall, independent of any known positioning mechanism such as the Min system~\cite{MinOscillations}. Other examples of  periodically positioned protein clusters have emerged as well~\cite{Janakiraman04, Lindner08}; for example, protein aggregates associated with cellular aging in bacteria exhibit a regular distribution along the the cell's long axis in filamentous {\it E. coli}~\cite{Lindner08}. The question arises -- could such periodic positioning arise spontaneously or does it require the existence of an unknown positioning system?

   Here we demonstrate, within the context of a minimal lattice model,
that protein clustering and periodic positioning of clusters can emerge spontaneously in growing cells. 
Lattice models have been used before to study clustering of membrane proteins with short-range interactions~\cite{Goldman04}. In our model,
 existing clusters act as sinks for proteins newly inserted in the membrane, locally reducing the density of protomers and thus preventing nucleation of new clusters. As cells grow, existing clusters separate, ultimately allowing new clusters to nucleate at a characteristic spatial separation set by insertion, diffusion, interaction strength, and growth rates. The proposed mechanism is quite general; while we focus on membrane proteins, the mechanism also applies to aggregation of cytoplasmic proteins in the body of the cell ({\it e.g.} misfolded protein aggregates)~\cite{Lindner08}.

\begin{figure}[htbp]
\begin{center}
\scalebox{0.25}{\includegraphics{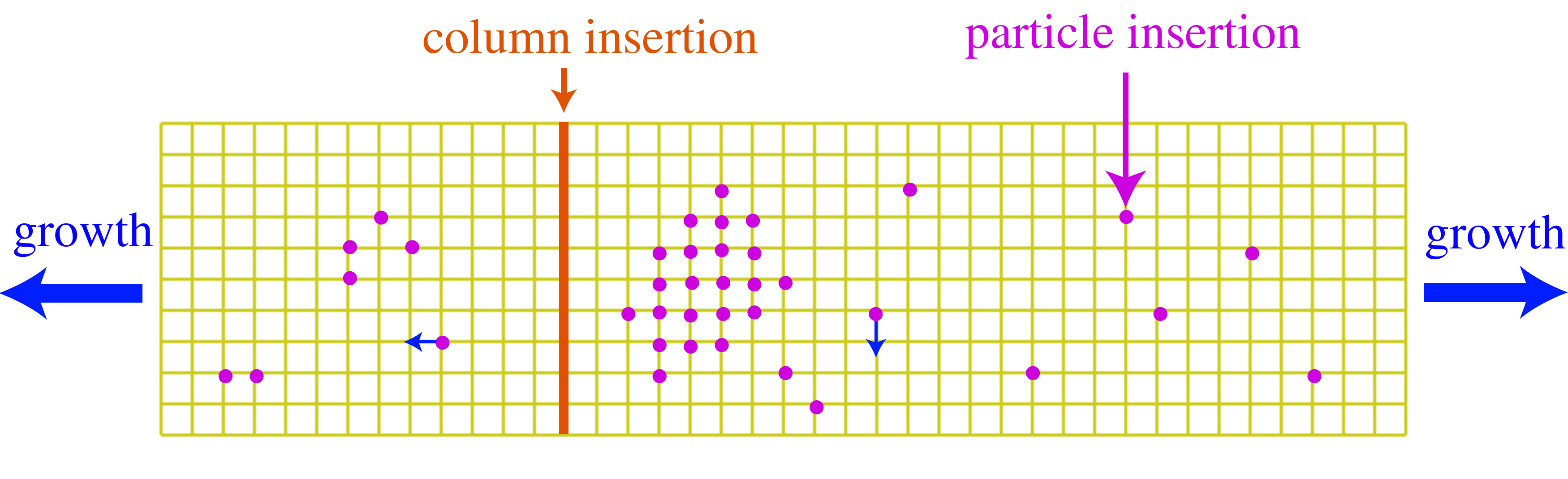}}
\caption{Schematic of the lattice model. Particles hop at random between neighboring lattice points and can join or leave an existing cluster from the boundary of the cluster. Columns of lattice points are inserted at random to mimic cell growth, and particles are inserted at random to mimic protein insertion in the membrane. }
\label{fig: chemo_schematic}
\end{center}
\end{figure}

 In our model, the cell membrane is represented by a square lattice, whose $x$-axis coincides with the long axis of the cell (see Fig.~\ref{fig: chemo_schematic}).  We employ  periodic boundary conditions in the $y$ direction to account for the cylindrical shape of bacteria like {\it E. coli}. The protomers   (independently diffusing protein units) associated with the cell membrane~\cite{Protomers}  are treated as particles which can perform random walks on the lattice. Each lattice site is therefore associated with a variable $\sigma_i$, either occupied, $\sigma_i = 1$, or empty, $\sigma_i = 0$. We assume a nearest-neighbor attractive interaction between particles with interaction energy $J$, measured in units of  the thermal energy $\kb T$.  To control the nucleation barrier, we also include a conformational energy cost given by $\am J$ 
for each particle with any neighbors, which accounts for the loss of internal entropy when a particle associates with a cluster or a second  protomer.
The total energy of the system (in units of  $\kb T$) is 
\begin{equation} 
E =  - J  \sum_{<i,j>}\sigma_{i} \sigma_{j}  \  + \ \am J \nm,
\label{hamiltonian}
\end{equation}
where $\nm$ is the total number of particles with one or more neighbors ({\it i.e.}, the number of particles that are in clusters of size two or greater).  
Experiments indicate that the lateral receptor clusters are relatively immobile while individual 
membrane proteins are typically free to diffuse~\cite{ThiemPositioningofchemosensoryEJ07}. In our lattice model, we therefore consider only movement of individual particles. 

 We use a Metropolis Monte Carlo algorithm to simulate the system. A randomly selected particle is  
moved to one of its unoccupied neighboring sites with an acceptance probability  $p = \min(1,e^{- \Delta E })$ where $\Delta E$ is the energy change due to the proposed displacement of the particle. 
One Monte Carlo time step corresponds to one attempted move for each particle present.  

 

In the absence of the conformational energy cost  ($\am =0$), the thermodynamic system described by our energy function 
can be mapped to a two-dimensional Ising model,  for which the critical interaction strength is known to be
 $\Jc  \approx 1.763$~\cite{Onsager44}. When the interaction strength is low, $J < \Jc$, the system has one stable homogeneous phase, while for $J > \Jc$, the system can phase separate into regions of high and low density. The conformational energy cost increases $J_{c}$. 

To account for growth of the bacterial cell, we allow the lattice to expand in the $x$ direction according to $\lx (t) \approx \lo e^{\gamma t}$, where $\lo$ is the initial length of the bacterium and $\gamma$ is the  growth rate. The expansion of the lattice is implemented by random insertion of empty columns  at a rate $\gamma \lx(t)$ with equal probability anywhere in the lattice.  Based on the observation that newly synthesized chemotaxis receptors are inserted into the cell membrane uniformly over the entire length of the cell~\cite{Shiomi06}, particles are randomly deposited onto the lattice at a rate $\kon$ per available ({\it i.e.} unoccupied) site. This ultimately leads to an average density of occupied sites $\Co =  \kon /(\gamma + \kon)$. To see this, 
let $N(t)$ and $n(t)$ be the total number of lattice sites 
and the number of occupied sites, respectively, at time $t$, with $N(t) = N(0) e^{\gamma t}$. On average, the total rate of particle deposition is given by $dn / dt = \kon [N(t) - n(t)]$; the general solution to this equation is given by $n(t) = \Co N(t) + C e^{- \kon t}$,  where $\Co =  \kon /(\gamma + \kon)$ is the asymptotic value of the  particle number density,  defined by $\rho(t) \equiv n(t) / N(t)$,
and $C$ is a constant.  We start our simulations with 
$\rho(t=0) = \Co$ so that, on average, $\rho (t)$ remains fixed at  $\Co$.

      For our simulations, the cell circumference was fixed at $\ly = 50$, with $\am = 0.5$ and interaction strength $J = 4$, considerably greater than the critical strength $\Jc$.
The system was initialized with a cluster at each end of the cell to mimic the existing clusters at the poles.  At the start of each simulation, the length of the cell is $\lo = 20$ and, as the cell grows, newly inserted particles aggregate to form clusters that grow with time. Fig.~2 illustrates a series of snapshots from a representative run with the growth rate  chosen to be $\gamma = 0.8 \times 10^{-5}$, and a deposition rate $\kon = 2 \times 10^{-6}$,  yielding $\Co = 0.2$. Notice that clusters of particles spontaneously appear at positions approximately periodically spaced along the cell.

\begin{figure}[htbp]
\begin{center}
\scalebox{0.3}{\includegraphics{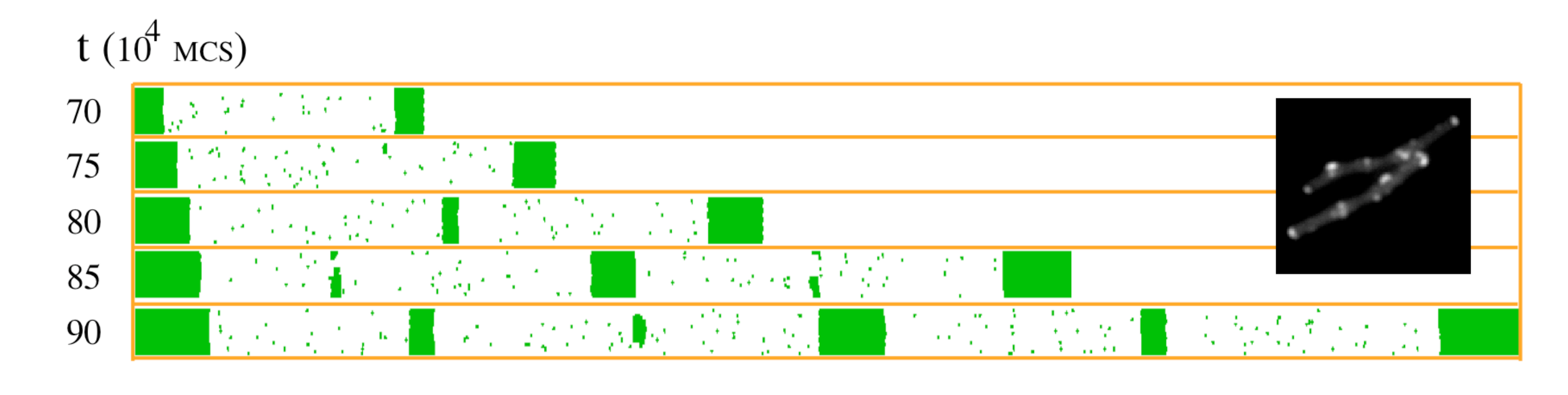}}
\caption{Snapshots of the model cell membrane at times 70, 75, 80, 85, and 90 $\times 10^{4}$
Monte Carlo time-steps as defined in the text. The total size of the system in the final snapshot is 
1350 columns by 50 rows, with $\approx 20 \%$ of lattice sites occupied by protein particles (green). In the inset we show a representative image of protein receptor clusters in {\it E. coli} from Victor Sourjik's lab (courtesy:  Victor Sourjik)}
\label{fig:chemo_snaps}
\end{center}
\end{figure}

    The emergence of a self-organized periodicity of clusters can be understood by noting the positions of new clusters when they first appear.  
At the start of the simulation the clusters at each end of the cell act as sinks for newly inserted particles.  As the cell grows and these two clusters move apart,
a  new cluster forms roughly at the midpoint of the cell. As cell growth continues, newly inserted particles spontaneously aggregate to form new clusters in between existing clusters; the location of a newly formed cluster is preferentially at the middle of two existing clusters, resulting in periodically positioned clusters (Fig.~\ref{fig:chemo_snaps}).
If the separation between two existing clusters is below a characteristic length $\lc$, diffusion dominates 
and particles are absorbed by old clusters;
if the separation is larger than $\lc$,  particles nucleate to form a new cluster, leading to periodic spacing $\sim \lc$. 

\begin{figure}[htbp]
\begin{center}
\scalebox{0.3}{\includegraphics[angle = 90]{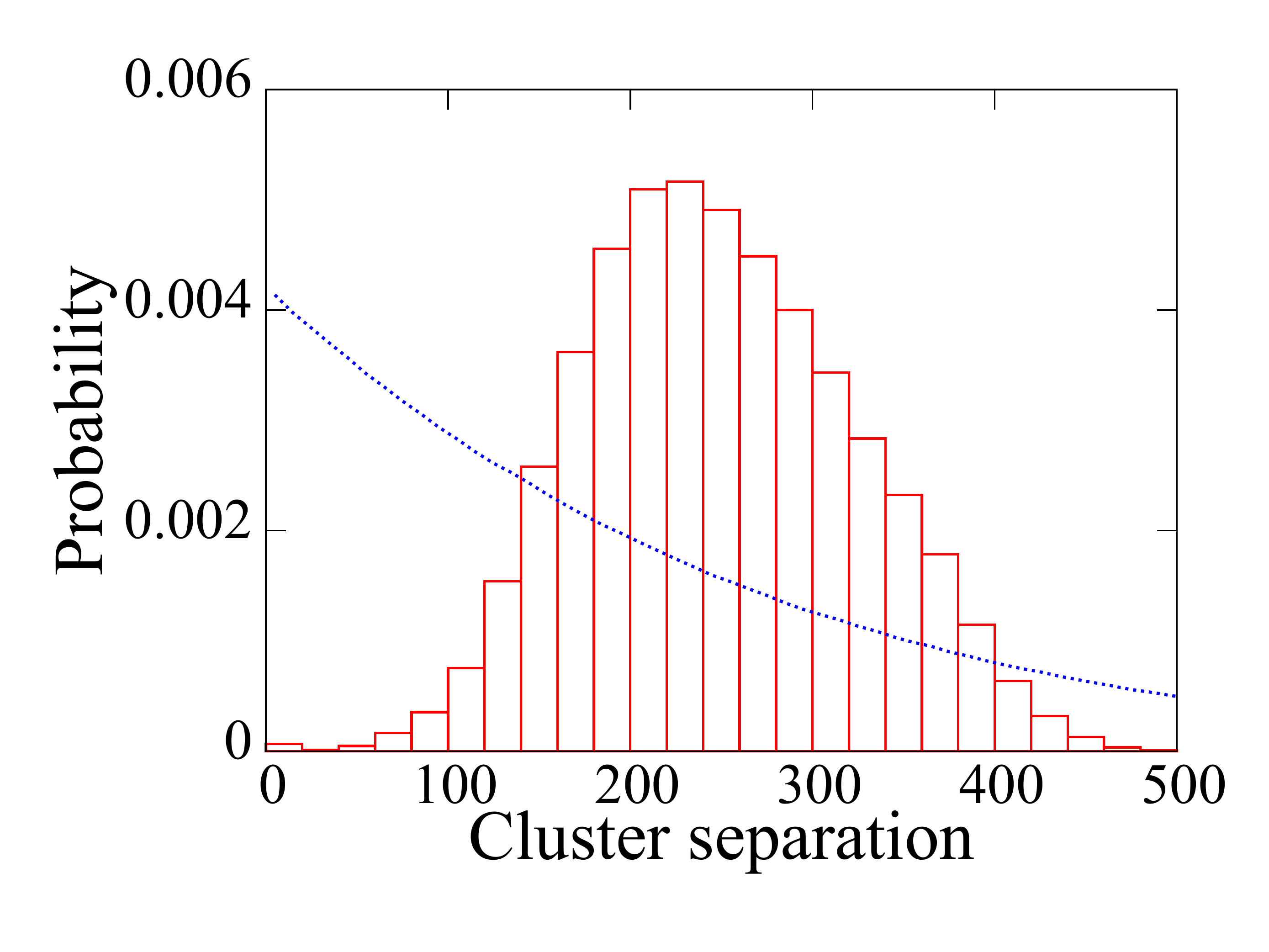}}
\caption{The distribution of inter-cluster separations. Simulations were stopped when the system reached $\lx = 1900$, which corresponds to approximately eight clusters in the system. The density is $\Co = 0.1$ and the growth rate is $\gamma = 1\times10^{-5}$. The data was averaged over 70,000 simulations. The bin size is 20. If the centers of clusters were randomly distributed, the distribution of inter-cluster separations would be exponential as shown by the dotted curve.}
\label{fig:fig3}
\end{center}
\end{figure}

To quantify the separation between clusters, we obtained the distribution of the separations between neighboring clusters for systems grown to $\lx = 1900$ 
(see Fig.~\ref{fig:fig3}).  Due to stochastic fluctuations, a cluster is not thermodynamically stable until it reaches a critical size. We found that clusters of size $\ge 50$ were stable and did not disappear; we thus 
 used size 50 as a criterion to identify a  cluster. The separation between neighboring clusters is defined to be the distance between the centers of mass of the two clusters. 
 For $\lx = 1900$, there are on average eight clusters present in the system. The distribution of separations exhibits a single maximum at $\Delta x \approx 230$, indicating a preferred separation between neighboring clusters. Moreover, the fraction of cluster separations less than half the peak value is only $7.6 \%$, indicating a strong suppression of close ({\it i.e.} $\Delta x < 115$)  clusters.   For comparison, the distribution of inter-cluster separation would be an exponential if the cluster centers were positioned randomly (dotted line in Fig.~\ref{fig:fig3}).

To investigate the mechanism responsible for cluster positioning, we studied how the position of a newly formed cluster depends on the positions of existing clusters. For simplicity, we used periodic boundary conditions in the $x$ direction and initialized simulations with $\lo = 1$, that is, with only one column of sites in the system. We used $\am = 0.5$, $J = 4.0$, $\kon =1.1 \times 10^{-6}$, and 
$\gamma = 10^{-5}$, yielding $\rho_{0} \approx 0.1$. 
As the system grows, the deposited  particles  aggregate to form first one cluster and, later, a second stable cluster. We recorded the position of the second cluster (once it had reached a size $\ge 50$) with respect to the first.  The  distribution of the separations between the new cluster and the existing cluster as a fraction of the total cell length is plotted in Fig.~4(a). The distribution peaks at a reduced distance of 0.5, {\it i.e.}, the second cluster is most likely to form at the midpoint, equidistant from the edges of the existing cluster.

To understand why the second cluster formed near midcell, 
we investigated the density profile of particles in the dilute region between two stable clusters during cell growth. Simulations were performed, as for Fig.~4(a), using periodic boundary conditions in $x$ and starting from a single column. In Fig.~4(b), we plotted the particle density profile along the $x$ axis as measured when the system reaches $\lx = 300$ at an average global particle density $\Co = 0.1$. In these simulations, at $\lx =300$, typically there was one stable cluster in the system. Position was measured from one edge of the cluster and  was normalized by the distance between the two flanking edges of the cluster. The average particle density fits very well to a quadratic function with the maximum at midcell. 

 \begin{figure}[htbp]
\begin{center}
\scalebox{0.22}{\includegraphics[angle = 0]{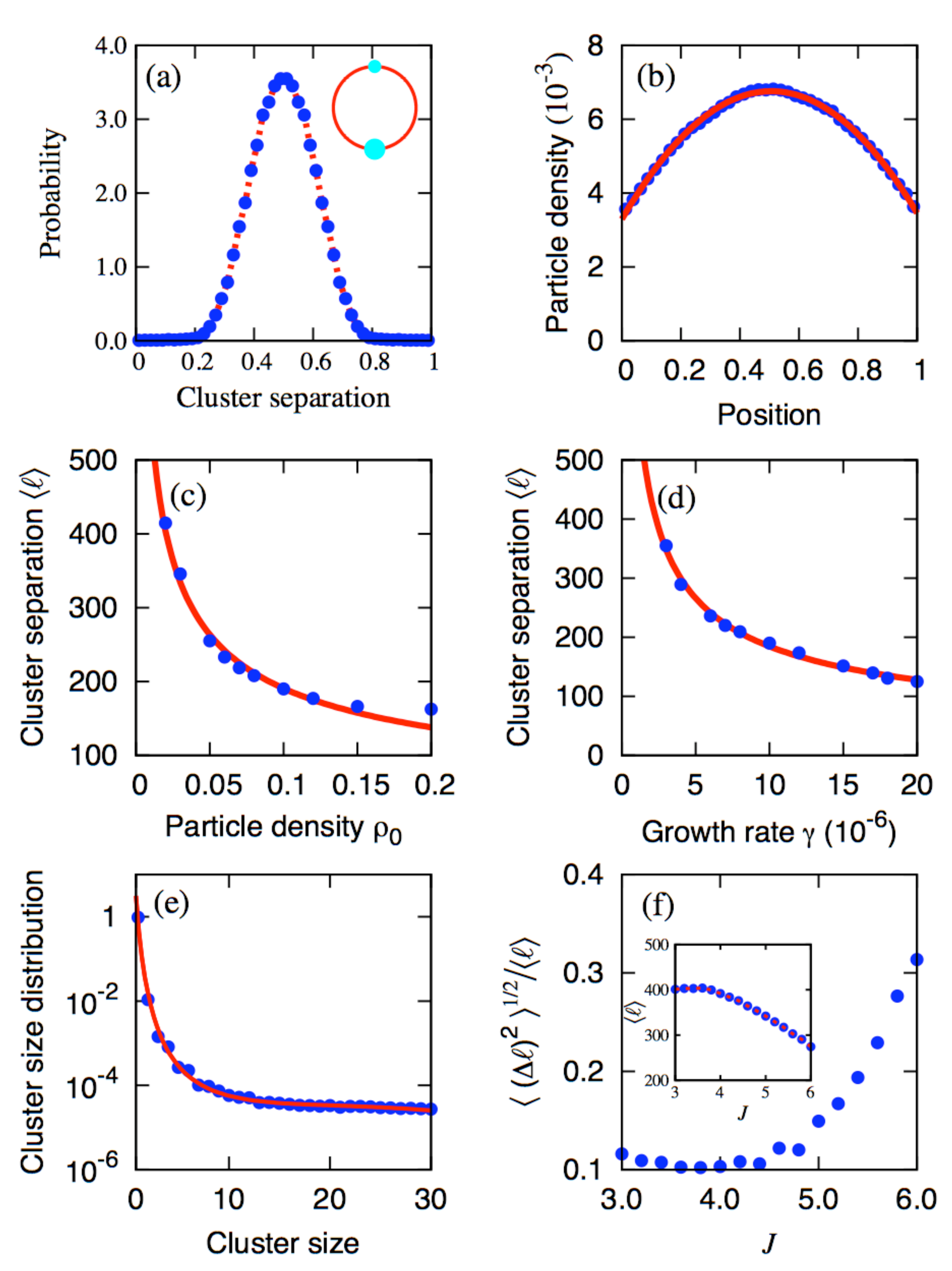}}
\caption{ (a) Distribution of the separation along the $x$ direction between the center of the new (second) cluster and the center of the old (first) cluster, using periodic boundary conditions in $x$ (see inset). The  dotted line is a guide to the eye. 
(b) The particle density profile in the $x$ direction (periodic boundary conds.) with one cluster in the system. 
The particle density was measured when the system reached $\lx = 300$ and was averaged over 10,000 simulations. The distance  was normalized by the separation between the two flanking edges of the cluster as measured in the $x$ direction. The smooth line is a fit to a parabola. (c) The average inter-cluster separation versus the particle density $\Co$ at fixed growth rate $\gamma = 10^{-5}$. (d) The average inter-cluster separation versus growth rate $\gamma$ at fixed particle density $\Co = 0.1$. 
The smooth curves  in (c) and (d) are fits to power  laws $\Co^{-0.47}$ and $\gamma^{-0.53}$ respectively.
(e) The cluster size distribution averaged over 10,000 simulations.  Cluster sizes were measured when the system reached $\lx = 1900$; 
the red line is a fit to the curve $y = \exp[a_{1}  x  + a_{2} x \ln x + a_{3} \ln x + a_{4}] $, with $a_{1} \approx 0.8$, $a_{2} \approx -0.37$, $a_{3} \approx -7.6$, and $a_{4} \approx -0.9$, for cluster size.       
 (f) Dependence of the standard deviation of cluster separation (normalized by the average separation) on  interaction strength 
 $J$.  The parameters and the method for collecting data were the same as in (a).  In the inset we show the dependence of inter-cluster separation on $J$.   }
\label{fig:fig4}
\end{center}
\end{figure}

The observed quadratic particle density profile can be understood as follows. 
The average local particle density $\rho({\bf r}, t)$ in regions that do not contain clusters satisfies the diffusion equation $ \partial \rho/ \partial t =D \nabla^{2} \rho + \kon$,
where $D$ is the particle diffusion coefficient and $\kon$ is the particle insertion rate.
Consider a region flanked by two stable neighboring clusters with flanking edges at $x = 0$ and $x = \ell$. Approximating stable clusters as perfect sinks for particles gives the boundary conditions $\rho(0) = \rho(\ell) = 0$. Hence, in the membrane strip between two clusters, the steady-state 
solution~\cite{footnote1} to the diffusion equation is 
\begin{equation}
\rho(x) = -\frac{\kon}{2 D}\left(x - \frac{\ell}{2}\right)^{2} + \frac{\kon \ell^{2}}{8 D}.
\label{eq:parabola}
\end{equation}
The peak of $\rho(x)$ is located at $x=\ell/2$, precisely at the midpoint between
 two cluster edges. The maximum particle density is $\rho_{\rm max} = \kon \ell^{2}/8 D$. Notice that it  scales quadratically with cluster spacing.


     Since the particle density peaks at the midpoint between two neighboring clusters, this is the most likely  place for a new cluster to nucleate. 
Of course, nucleation of a new cluster is a stochastic event;
nevertheless,  the probability of nucleating a new cluster is a highly nonlinear function of the local density, so the quadratic density profile gives rise to the sharply peaked distribution for the position of the new stable cluster  (Fig.~4(a)).    

    The existence of a relatively sharp density threshold for cluster nucleation predicts scaling relations for the mean cluster separation. For steady-state growth of a long cell, on average a new cluster must appear between neighboring old clusters every time the cell doubles ({\it e.g.}, see Fig.~\ref{fig:chemo_snaps}). If there is a sharp density threshold $\Cth$ for cluster nucleation, then a new cluster will nucleate  when the peak density between old clusters reaches approximately this value. We can therefore estimate the upper limit of cluster separation in terms of $\Cth$ (with the lower limit being a factor of two smaller, and the mean lying between the two.) According to Eq.~\ref{eq:parabola}, the peak density between clusters depends on their separation $\ell$ according to $\rho_{\rm max} = \kon \ell^{2}/8 D$, where $\kon = \rho_o \gamma/(1-\rho_o) \simeq \rho_o \gamma$. Nucleation of a new cluster should therefore occur when $\rho_{\rm max} \approx \Cth$, that is, for a cluster separation $\lc \approx 2[2D\Cth/(\rho_o \gamma)]^{1/2}$. This implies that the mean cluster separation in a growing cell will obey the scaling relations $\lc \sim\rho_{0}^{-1/2}$ and $\lc  \sim\gamma^{-1/2}$ \cite{footnote2}, which are verified in Figs.~4(c) and 4(d).

         We have demonstrated that an interplay of protein creation or insertion in the membrane, protomer diffusion, aggregation,  and cell growth can lead to periodically spaced protein clusters. In addition to the larger stable clusters, our stochastic nucleation mechanism also generates a 
 quasi-steady state distribution of smaller clusters as shown in Fig.~4(e). Finally, in Fig.~4(f), we plot the dependence of the standard deviation of inter-cluster separation on the interaction strength $J$; the  
data implies that the periodic placement of protein clusters is robust to variations in $J$ provided $J < 5$.

    For chemoreceptors in {\it E. coli}, the observed inter-cluster spacing is of the order of $ 1 \mu$m. Assuming a  membrane diffusion constant of $D = 0.018 \mu$m$^{2}$/sec~\cite{Diffusion_constant} and doubling time of around 60 minutes,  we estimate $ \Cth \approx 2.4 \times 10^{-3} \Co $, implying that the vast majority of the chemoreceptors are bound to clusters rather than existing as free protomers. This prediction is testable using modern fluorescent imaging techniques~\cite{PALM}.
 In principle, the scaling relations for $\lc$ as a function of receptor density and growth rate can also be tested experimentally by measuring cluster spacings in cells over- or under-producing chemoreceptors and growing at different rates. Moreover, our model predicts that if proteins are produced (and hence inserted into the membrane) in bursts rather than continuously, it would adversely affect the periodic placement of clusters.


     Finally, we consider the biological significance of the periodicity of protein clusters. In the case of chemoreceptors, periodic cluster spacing ensures that each daughter cell receives at least one cluster following cell division or fragmentation of a filamentous cell~\cite{ThiemPositioningofchemosensoryEJ07}. 
This ``equipartition'' of protein clusters among daughter cells might be advantageous as well for other protein complexes. In contrast, the larger spacing between misfolded protein aggregates typically means that only one daughter receives an aggregate; this unequal partition results in different fitnesses of the two daughters, with a potential overall reproductive advantage for the population~\cite{Lindner08}.


 We thank K.C. Huang, Y. Meir, and V. Sourjik for helpful discussions and suggestions. NSW and RM acknowledge 
 support from NIH grant R01 GM073186.


\begin{thebibliography}{99}
\bibitem{footnote1}
For the quasi-steady state solution to accurately describe the actual density profile in the growing system, the doubling time has to be much larger than the time for a particle to diffuse over a length-scale comparable to its size (or lattice spacing, in the simulations). This is satisfied both in our simulations and for the real bacterial cell.
\bibitem{footnote2}
In principle, the relevant nucleation threshold, $\Cth$, also depends on the doubling time, but only very weakly -- the rate of nucleating a new cluster is $\sim \rho_{\rm max}^{N_c}$, where $N_c$ is the critical nucleus size, so a twofold increase in the doubling time, say, will only change $\Cth$ by  $(1/2)^{1/N_{c}} $.
\end{thebibliography}
\end{document}